\documentclass[conference, letterpaper]{IEEEtran}
\IEEEoverridecommandlockouts
\usepackage{cuted}
\usepackage{cite}
\usepackage{bm}
\usepackage{amsmath,amssymb,amsfonts}
\usepackage{algorithm}
\usepackage{graphicx}

\usepackage{algpseudocode}

\usepackage{graphicx}
\usepackage{subcaption}

\usepackage[letterpaper, left=0.625in, right=0.625in, bottom=1.02in, top=0.70in]{geometry}

\usepackage{textcomp}
\usepackage{xcolor}

\newcommand{\be}{\begin{equation}}
\newcommand{\ee}{\end{equation}}

\def\BibTeX{{\rm B\kern-.05em{\sc i\kern-.025em b}\kern-.08em
    T\kern-.1667em\lower.7ex\hbox{E}\kern-.125emX}}

\begin{document}

\title{Energy Efficient Robust Beamforming for Vehicular ISAC   with Imperfect Channel
Estimation}
\author{\IEEEauthorblockN{Hanwen Zhang\IEEEauthorrefmark{1}, Haijian Sun\IEEEauthorrefmark{1}, Tianyi He\IEEEauthorrefmark{2}, Weiming Xiang\IEEEauthorrefmark{3}, Rose Qingyang Hu\IEEEauthorrefmark{2}} 
\IEEEauthorblockA{
\IEEEauthorrefmark{1}School of Electrical and Computer Engineering, University of Georgia, Athens, GA, USA \\
\IEEEauthorrefmark{2}College of Engineering, Utah State University, Logan, UT, USA \\
\IEEEauthorrefmark{3}School of Computer and Cyber Sciences, Augusta University, Augusta, GA, USA \\
Emails: hanwen.zhang@uga.edu, hsun@uga.edu, tianyi.he@usu.edu, wxiang@augusta.edu, rose.hu@usu.edu}}
\maketitle

\begin{abstract}
This paper investigates robust beamforming for system-centric energy efficiency (EE) optimization in the vehicular integrated sensing and communication (ISAC) system, where the mobility of vehicles poses significant challenges to channel estimation. To obtain the optimal beamforming  under channel  uncertainty, we first formulate an optimization problem for maximizing the system EE under bounded channel estimation errors. Next, fractional programming and semidefinite relaxation (SDR) are utilized to relax the rank-1 constraints. We further use Schur complement and \emph{S}-Procedure to transform Cram\'er-Rao bound (CRB) and channel estimation error constraints into convex forms, respectively. Based on the Lagrangian dual function and Karush-Kuhn-Tucker (KKT) conditions, it is proved that the optimal beamforming solution is rank-1. Finally, we present comprehensive simulation results to demonstrate two key findings: 1) the proposed algorithm exhibits a favorable convergence rate, and 2) the approach effectively mitigates the impact of channel estimation errors.

\end{abstract}


\section{Introduction}
There has been a growing demand to enable vehicle-to-everything (V2X) technologies for future autonomous vehicles. Empowered by advanced edge computing, sensors, and control algorithms, V2X can facilitate real-time decision-making, seamless perception, and critical alerts for vehicles \cite{10061429}. Among various V2X technologies, dedicated short-range communication \cite{DSRC_Pc} and cellular V2X \cite{cv2x} stand out as two representative efforts from the past few decades.  However, V2X alone cannot fulfill the requirements for complete vehicle intelligence as it lacks real-time sensing capabilities. To address this challenge, integrated sensing and communications (ISAC) has emerged as a promising solution that integrates both communication and sensing functions from the same wireless signals \cite{IoTJ_review_vcn,Network_V2X_review,du2023towards}. ISAC has gained significant attentions recently \cite{10251151,9416177,9652071,9761984,zou2023energy,Network_V2X_review}, and multiple studies have demonstrated its ability to enhance spectrum efficiency, enable target detection, achieve high tracking accuracy, and increase throughput \cite{10251151,cv2x,9652071,9761984,li2023isac9652071}.

There is a trade-off between communication and sensing functions in the ISAC system. Depending on the system design objectives, ISAC can be categorized as sensing-centric, communication-centric, or a combination of both in a joint design. Generally, the signal-to-interference-plus-noise ratio (SINR) is considered as a performance metric for communication, while the Cram\'er-Rao bound (CRB) is used for assessing sensing performance. As an example, in \cite{9652071}, the authors presented a framework for a multi-user multi-input-multi-output (MU-MIMO) sensing-centric ISAC system, with the objective of maximizing CRB performance.  In \cite{10217169}, the authors extended these results and conducted CRB analysis under both a simplified point target model and a more realistic extended target model.  On the other hand, \cite{10009894} proposed a communication-centric energy efficiency (EE) objective in ISAC and applied both SINR and CRB as design constraints. 

However, the aforementioned studies are not directly applicable to the V2X ISAC scenario considered here. The primary reason is that vehicle mobility poses challenges in acquiring accurate channel state information (CSI) \cite{IoTJ_review_vcn,Network_V2X_review,du2023towards}. Imperfect CSI has been considered mainly in  communication-centric wireless systems. To name a few, 
in \cite{9889690}, authors applied first-order Taylor approximation to address the unstable channel estimation errors. In \cite{8479337,7922522}, authors provided robust algorithms to tackle the bounded unstable channel estimation factors in non-orthogonal multiple access systems. By applying the \emph{S}-procedure, they reformulated the optimization problem with channel uncertainty factors into a convex form.
In the ISAC system, imperfect CSI can complicate SINR and CRB modeling, as well as the analysis of their trade-offs.

In this work, we introduce channel uncertainty into the ISAC system, where the error in CSI estimation is constrained within certain bounds. For the communication part, we select system EE, which represents the sum EE of all vehicle users, as the design objective. EE offers an optimal trade-off between achievable data rates and power consumption, making it a popular consideration in green communication design. For the sensing part, we use CRB as the performance metric. The major contributions of the paper are summarized as follows: 
\begin{itemize}

\item We formulate an optimization problem with the objective of maximizing system EE while accounting for CSI estimation errors. The problem is challenging to solve due to the nature of the sum-of-ratio objective function and the presence of bounded channel error models. To address that, we employ fractional programming to reframe the objective function, utilize semidefinite relaxation (SDR) to relax the rank-1 constraint, and apply the \emph{S}-procedure to address channel uncertainty in the SINR constraint. The CRB constraint is further transformed into semidefinite form using the Schur complement. As a result, the problem has been successfully converted to convex.

\item To obtain the optimized robust beamformer, we have introduced an iterative algorithm.  Based on the initialized fixed parameters, the original problem is optimized with semidefinite beamforming result. We then utilize eigenvalue decomposition (EVD) to obtain the beamforming vectors. To address channel uncertainty in the fixed parameters, we integrate the optimal results from the original problem and formulate a sub-problem to update the fixed parameters sequentially.

\item By using Lagrangian dual function and Karush-Kuhn-Tucker (KKT) conditions, we prove that the rank of beamforming results in this problem is 1 with matrix form CSI. This justifies the use of SDR and EVD.

\item Simulation results have validated the proposed algorithm. We have shown that the algorithm has a good convergence rate, and the proposed solution can effectively address channel uncertainty. 
\end{itemize}

The remainder of the paper is organized as follows. The system model is introduced in Section II, encompassing the communication, sensing, and EE design aspects. Section III provides the robust optimization problem formulation. In Section IV, we present the solutions, which involve the application of the \emph{S}-procedure, Schur complement, and SDR. Additionally, we provide details on the iterative algorithm and the proof of rank-1 beamforming. The simulation results are discussed in Section V. Section VI concludes the paper. 
\section{System Model}
\begin{figure}[!h]
    \centering
    \includegraphics[width=2.7in]{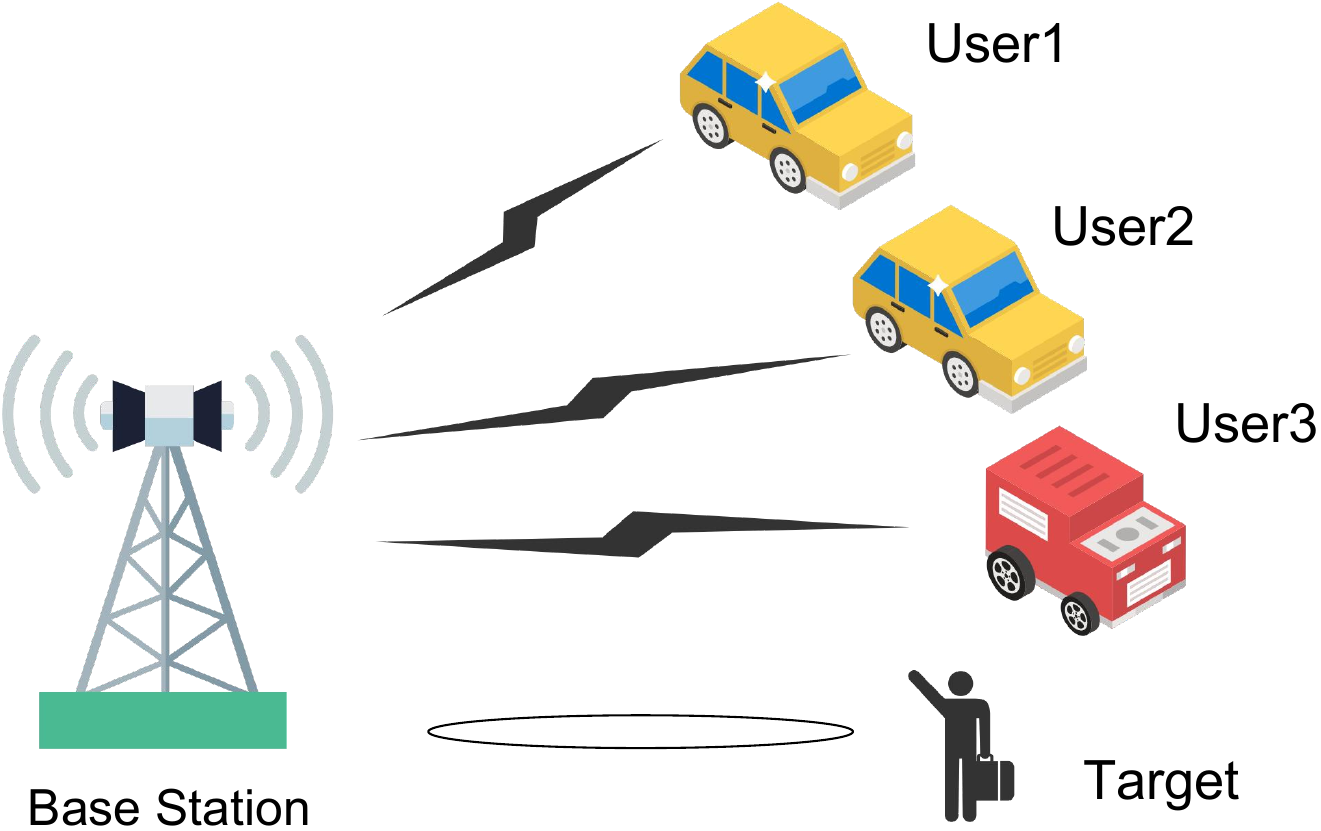}
	\centering
	\caption{System architect}
	\label{fig:model}
\end{figure}

As shown in Fig. \ref{fig:model}, The system comprises a monostatic ISAC base station (BS) operating in the high-frequency band and equipped with an antenna array consisting of $M$ elements to compensate for the high pathloss. Out of these $M$ elements, $M_t$ elements are allocated for data communication with $K$ users, while $M_r$ elements are exclusively reserved for sensing purposes, i.e. $M = M_t + M_r$, with $M_t < M_r$. The considered model that separates communication and sensing antennas proves to be effective in tackling excessive self-interference.
At the received side, 
each vehicle user has $N$ antenna elements for better signal detection and transmission. 
\subsection{Communication Model}
To take advantage of the antennas array, BS generates beamforming vectors for each user. The baseband signal is denoted as $\mathbf{x}[n] = \sum_{k=1}^{K} \mathbf{w}_k s_k [n], \forall k \in \mathcal{K}$, $\mathcal{K}=\{ 1,2,3...,K\}$, where $\mathbf{w}_k \in \mathbb{C}^{M_t \times 1}$ is the beamforming vector for the $k$-th user. $s_k[n]$ is the data symbol with unit power $E(|s_k[n]|^2) = 1$, $n$ is the discrete time index. At the receiver end, user $k$ receives $\mathbf{y}_k [n]$ at time $n$, denoted as
\be 
\mathbf{y}_k [n] = \mathbf{H}_k \mathbf{x}[n] + \mathbf{n}_m[n],\label{eq1}
\ee

where $\mathbf{n}_m[n]$ is the additive noise that follows $\mathbf{n}_m[n] \sim \mathcal{CN} (0, \sigma_m^2\mathbf{I})$. Different from existing works that assume $\mathbf{H}_k \in \mathbb{C}^{N\times M_t}$ can be perfectly estimated, the considered system takes channel estimation error into consideration. Specifically, let $\mathbf{H}_k = \hat{\mathbf{H}}_k + \Delta \mathbf{H}_k$, where $\hat{\mathbf{H}}_k$ is the estimation and $\Delta \mathbf{H}_k$ is the corresponding estimation error. We model the imperfect channel estimation under a bounded error case, where the spectral-norm of the error matrix are confined to a maximum value, i.e., 
\be 
\mathbf{\Phi}_k \triangleq \big\{   \Delta  \mathbf{H}_k \in \mathbb{C}^{N \times M_t} : ||\Delta  \mathbf{H}_k|| \leq  \phi   \big\}, \ \forall k,  \label{eq3}
\ee
where $||.||$ represents spectral-norm. This channel estimation error model has been applied extensively in literature \cite{spectral_norm_research}. 
The achievable SINR of user $k$ is expressed as 
\be \text{SINR}_k =   {\mathbf{w}_k^H\mathbf{H}_k^H }
(\sum_{i=1, i \neq k }^K  \mathbf{H}_k\mathbf{w}_i\mathbf{w}_i^H\mathbf{H}_k^H + \sigma_m^2\mathbf{I})^{-1}{\mathbf{H}_k\mathbf{w}_k }. \label{eq2}
\ee
Based on (\ref{eq2}), the achievable data rate for each user becomes 
\be
R_k = \text{B} \text{log}_2 (1+ \text{SINR}_k), \label{eq4}
\ee
where $\text{B}$ is the bandwidth. Without loss of generalizability, let $\text{B}=1$ in following sections.

\subsection{Sensing Model}
The transmitted signal $\mathbf{x}[n]$ can also be used for sensing activity. Essentially, electromagnetic waves of data signal $\mathbf{x}$ should be reflected back to sensing antennas when encountering objects. In general, $\mathbf{x}[n]$ is a pulse-shaping, narrowband signal. To improve sensing accuracy, a sequence of messages $\mathbf{X} = \{ \mathbf{x}[1], \mathbf{x}[2], \ldots, \mathbf{x}[L]\} \in \mathbb{C}^{M_t \times L}$ that spanning in $L$ time series are used for detecting objects. Therefore, the reflected signal can be expressed as 
\be 
\mathbf{Y}_s = \mathbf{G}_s \mathbf{X} + \mathbf{N}_s,\label{eq5}
\ee 
where $\mathbf{N}_s \sim \mathcal{CN} (0, \sigma_s^2 \mathbf{I}) $ is the additive noise, $\mathbf{G}_s$ is the aggregated channel response from the transmitter to the target then to the receiver array. Depending on the target type, $\mathbf{G}_s$ can be modeled in different formats. Overall,  we assume that targets are in the line-of-sight (LoS) and are situated in the far-field of the BS. Thus, it can be treated as a single-point \cite{9652071}. Let $\theta$ be the relative azimuth angle from the BS to the target. With LoS setting, the sensing channel response is 
\be 
\mathbf{G}_s = \alpha \mathbf{b}(\theta) \mathbf{a}^H (\theta),\label{eq6}
\ee 
\begin{strip}
\be 
\text{CRB}(\theta) = \frac{\sigma_s^2 \text{Tr}(\mathbf{A}^H(\theta)\mathbf{A}(\theta) \mathbf{R}_x)}{2 |\alpha|^2 L (\text{Tr}(\mathbf{\dot{A}}^H(\theta)\mathbf{\dot{A}}(\theta) \mathbf{R}_x) \text{Tr}(\mathbf{A}^H(\theta)\mathbf{A}(\theta) \mathbf{R}_x) -|\text{Tr}(\mathbf{\dot{A}}^H(\theta)\mathbf{A}(\theta) \mathbf{R}_x)|^2 ) },\label{eq7}
\ee 
\be
{R}_{fp,k}={\text{log}_2(1+2\sqrt{\text{Re}\{{\mathbf{z}_k^H}{\mathbf{H}_k}{\mathbf{W}_k}{\mathbf{H}_k^H}{\mathbf{z}_k}\}}-\mathbf{z}_k^H({\sigma_m^2}\mathbf{I}+{\sum_{i=1, i \neq k }^K \mathbf{H}_k\mathbf{W}_i\mathbf{H}_k^H)\mathbf{z}_k)}}.\label{eq8}
\ee
\end{strip}
$\alpha$ accounts for the combination of path-loss, reflection coefficient, and other radar effects. $\mathbf{a} \in \mathbb{C}^{M_t\times 1}$ and $\mathbf{b} \in \mathbb{C}^{M_r\times 1}$ are steering vectors of the transmit and sensing antenna arrays, respectively. For notation simplicity, let $\mathbf{A(\theta)} = \mathbf{b}(\theta) \mathbf{a}^H(\theta)$.  

The objective of sensing is to estimate the target location. In our case, this is equivalent to estimating $\alpha$ and $\theta$ via the received signal $\mathbf{Y}_s$. We use CRB for $\theta$ as the sensing performance metric, as it represents the lower bound of the variance of all the unbiased estimators. Since this signal contains Gaussian noise term $\mathbf{N}_s$, the CRB for $\theta$ is given by \cite{9652071} and is shown as  (\ref{eq7}),
where $\mathbf{R}_x = \mathbb{E}(\mathbf{X} \mathbf{X}^H) = \sum_{k=1}^K \mathbf{w}_k \mathbf{w}_k^H$.

\subsection{Energy Efficient Communication}
To support energy efficient communication, we select sum of each user's EE as the performance metric. Specifically, system EE is defined as
\be
\text{EE} =\sum_{k=1}^K f_k\frac{ R_k}{P_k}. \label{eq9}
\ee
Here, $f_k$ is the weight for user $k$, and $\sum_{k=1}^K f_k=1$. Moreover, $P_k = \text{Tr}(\mathbf{w}_k \mathbf{w}_k^H) + P_{0}/K$, the $\text{Tr}(\mathbf{w}_k \mathbf{w}_k^H)$ is the transmission power for user $k$ while the $P_{0}/K$ is the fixed power consumption of user $k$, and $P_0$ is the total fixed power for the ISAC system. EE strikes a balance between achievable data rate and power consumption which has been regarded as a popular communication-centric metric.

\section{Robust Optimization Problem Formulation}
In this section, we formulate the optimization problem to maximize the system EE while ensuring both communication and sensing requirements being met.
\begin{subequations}
    \begin{flalign}
        \mathbf{P}_1 :&\max_{\mathbf{w}_k}   \text{EE} = \sum_{k=1}^K f_k\frac{ R_k}{P_k} \label{eq10a}\\
        \text{s.t.}\
        &\text{CRB}(\theta) \leq \rho, \label{eq10b}\\
        &\text{Tr}(\mathbf{w}_k \mathbf{w}_k^H) \leq P^{\text{max}}, \label{eq10c}\\
        &\Delta  \mathbf{h}_k \in  \mathbf{\Phi}_k, \ \forall k, \label{eq10d}\\
        &\text{SINR}_k\geq \zeta_k\label{eq10e}.
    \end{flalign}
\end{subequations}
The objective is to maximize the weighted system EE.  (\ref{eq10b}) limits the sensing boundary, which essentially ensures a satisfactory sensing (detection) performance by limiting CRB to be no more than $\rho$;  (\ref{eq10c}) gives the maximum  transmit power consumption. The channel estimation error is contained in  $ \mathbf{\Phi}_k, \forall k$, which is defined in  (\ref{eq3}). For the communication part, a minimum $\text{SINR}_k$ of $\zeta_k$ is required. The parameters to be optimized are the beamforming vector $\mathbf{w}_k$ for each user. 

This optimization problem is non-convex and hard to be solved directly. The difficulty comes from the following aspects: 1) the objective function is a fractional one with $\mathbf{w}_k$ in both nominator and denominator; 2) Similarly, the CRB function also has $\mathbf{w}_k$ contained in both its nominator and denominator; 3) The channel estimation error is implicitly contained in $R_k$. In the following section, we propose a series of transformations to simplify this optimization problem.

\section{Proposed Solutions}
In this section, we firstly address the non-convexity of the objective function using fractional programming \cite{8314727}.  Next, we tackle the issue of imperfect channels through the \emph{S}-procedure, followed by applying Schur complement to address the non-convexity of CRB.  After establishing the convex form of the original problem, we define a sub-problem for sequential parameter updates. Finally, we offer a sketch of proof to show that the optimal beamformers are rank-1.

\subsection{Fractional Programming for Objective Function  }
Fractional programming,  an iterative transformation of fractional terms, is applied  to the objective function in  (\ref{eq10a}). To tackle non-convexity caused by the coupled $\mathbf{w}_k\mathbf{w}_k^H$, let  $\mathbf{W}_k=\mathbf{w}_k\mathbf{w}_k^H$, $\mathbf{W}_k\in \mathbb{C}^{M_t \times M_t}$. Note that the equivalent transformation requires $\text{rank}(\mathbf{W}_k) = 1$. Based on \cite{8314727}, the objective function is equivalent to 
\be \text{EE} =   \sum_{k=1}^K f_k(2 q_k\sqrt{ {R}_{fp,k}}-q_k^2{P_\text{k}}), \label{eq11}\ee
where $R_{fp,k}$ is given by  (\ref{eq8}), $R_{fp,k}$ is derived from $R_k$. $q_k \in  \mathbb{R}$,  and $\mathbf{z}_k \in  \mathbb{C}^{N\times 1}$ are auxiliary variables. With fixed $\mathbf{W}_k$, the optimal value of $q_k$ and $\mathbf{z}_k$ can be calculated as 
\be q_k^*=\frac{\sqrt{R_k}}{\text{Tr}(\mathbf{W}_k)+P_0/K}\label{eq12},
\ee
\be \mathbf{z}_k^*={(\sigma_m^2\mathbf{I}+\sum_{i=1, i \neq k }{\mathbf{H}_k}{\mathbf{W}_i}\mathbf{H}_k^H)^{-1}\mathbf{H}_k\mathbf{w}_k}\label{eq13}.
\ee
Note that fractional programming transform  (\ref{eq10a}) into a convex form but requires the objective function and auxiliary variables to be updated alternatively.
\subsection{Channel Robustness and CRB Reformulation}
To tackle the non-convexity of constraint  (\ref{eq10d}), we define auxiliary variables $r_k$, $\epsilon_k$, and $v_k$.  $\mathbf{P}_1$ is then equivalently converted  to $\mathbf{P}_2$. $A_1$ is the set of variables, where $A_1= \{\mathbf{W}_k  ,r_k,\epsilon_k,v_k\}$
\begin{subequations}
    \label{eq14}
    \begin{flalign}
        \mathbf{P}_2:\hspace{1mm} &{\max_{A_1}} \hspace{1mm}\text{EE} =   {\sum_{k=1}^K f_k(2 q_k\sqrt{\text{log}_2(1+r_k)}}-q_k^2{P_k})\label{eq14a}\\
        \text{s.t.}\
        &\text{Tr}(\mathbf{W}_k) \leq P^{\text{max}}, \label{eq14c} \\
        &r_k\geq \zeta_k,\label{eq14e}\\
        &2\sqrt{\text{Re}\{ \epsilon_k\}}-v_k\geq r_k,\label{eq14f}\\
        &{\mathbf{z}_k^H}{\mathbf{H}_k}{\mathbf{W}_k}{\mathbf{H}_k^H}{\mathbf{z}_k}\geq \epsilon_k,\label{eq14g}\\
        &{\mathbf{z}_k^H(\sigma_m^2\mathbf{I}_{N}+\sum_{ i \neq k }^K \mathbf{H}_k\mathbf{W}_i\mathbf{H}_k^H)\mathbf{z}_k}\leq v_k\label{eq14h},\\
        &\text{(\ref{eq10b})},\text{(\ref{eq10d})}.\notag
    \end{flalign}
\end{subequations}
The $\mathbf{W}_k$ rank requirement is dropped per SDR. We substitute $\mathbf{H}_k=\hat{\mathbf{H}}_k+\Delta \mathbf{H}_k$ into  (\ref{eq14g}) and  (\ref{eq14h}) to take imperfect channel into consideration. Let $\mathbf{\xi}=\Delta \mathbf{H}_k^H \mathbf{z}_k$, $\mathbf{\Lambda}_k=\sum_{ i \neq k }^K \mathbf{W}_i$, The new expressions for  (\ref{eq14g}) and  (\ref{eq14h}) are given in  (\ref{eq15}) and  (\ref{eq16}) respectively. 
\begin{flalign}
&{\mathbf{z}_k^H}{\hat{\mathbf{H}}_k}{\mathbf{W}_k}{\hat{\mathbf{H}}_k^H}{\mathbf{z}_k}+
{\mathbf{\xi}^H}{\mathbf{W}_k}{\hat{\mathbf{H}}_k^H}{\mathbf{z}_k}+
{\mathbf{z}_k^H}{\hat{\mathbf{H}}_k}{\mathbf{W}_k}{\mathbf{\xi}}+{\mathbf{\xi}^H}{\mathbf{W}_k}\mathbf{\xi}\notag\\
&\geq \epsilon_k\label{eq15},\\
&\sigma_m^2{\mathbf{z}_k^H}{\mathbf{z}_k}+{\mathbf{z}_k^H}{\hat{\mathbf{H}}_k}\mathbf{\Lambda}_k{\hat{\mathbf{H}}_k^H}{\mathbf{z}_k}+
{\mathbf{\xi}^H}\mathbf{\Lambda}_k{\hat{\mathbf{H}}_k^H}{\mathbf{z}_k}+
{\mathbf{z}_k^H}{\hat{\mathbf{H}}_k}\mathbf{\Lambda}_k\mathbf{\xi}\notag\\
&+{\mathbf{\xi}^H}\mathbf{\Lambda}_k\mathbf{\xi}\leq v_k\label{eq16}.
\end{flalign}
With \emph{S}-procedure \cite{8479337}, we formulate  (\ref{eq15}) and  (\ref{eq16}) as the following semidefinite form:

\be
\begin{bmatrix}
{\mathbf{W}_k+{\lambda}_{1,k} \mathbf{I}_{M_t}}&{\mathbf{W}_k^H \hat{\mathbf{H}}_k^H \mathbf{z}_k}\\
{\mathbf{z}_k^H\hat{\mathbf{H}}_k \mathbf{W}_k }&{\text{B}-{\lambda}_{1,k}\phi_k^2 \mathbf{z}_k^H  \mathbf{z}_k}
\end{bmatrix}\succeq0, \label{eq18}
\ee

\be
\begin{bmatrix}
{-\mathbf{\Lambda}_k+{\lambda}_{2,k} \mathbf{I}_{M_t}}&{-\mathbf{\Lambda}_k^H \hat{\mathbf{H}}_k^H \mathbf{z}_k}\\
{-\mathbf{z}_k^H\hat{\mathbf{H}}_k \mathbf{\Lambda}_k }&{\text{C}-\mathbf{\lambda}_{2,k}\phi_k^2 \mathbf{z}_k^H  \mathbf{z}_k}
\end{bmatrix}\succeq0. \label{eq19}
\ee
Here, $\text{B}={\mathbf{z}_k^H}{\hat{\mathbf{H}}_k}{\mathbf{W}_k}{\hat{\mathbf{H}}_k^H}{\mathbf{z}_k}-\epsilon_k$, $\text{C}=-{\mathbf{z}_k^H}{\hat{\mathbf{H}}_k}{\mathbf{\Lambda}_k}{\hat{\mathbf{H}}_k^H}{\mathbf{z}_k}-\sigma_m^2{\mathbf{z}_k^H}{\mathbf{z}_k}+v_k$. $\lambda_{1,k}$, $\lambda_{2,k}$ are non-negative auxiliary variables.

To simplify  (\ref{eq10b}), we employ the Schur complement given by\cite{9652071} to represent CRB constraint as:
\be
{\text{Tr}(\mathbf{\dot{A}}^H(\theta)\mathbf{\dot{A}}(\theta)\mathbf{R}_x)-\text{D}-\Gamma\geq0}\label{eq20},
\ee
where the $\text{D}={|\mathbf{\dot{A}}^H(\theta)\mathbf{A}(\theta)\mathbf{R}_x|}^2\text{Tr}^{-1}(\mathbf{A}^H(\theta)\mathbf{\dot{A}}(\theta)\mathbf{R}_x)$ and $\Gamma={\sigma_s^2}/{(2{|\alpha|}^2L\rho)}$. Then we reformulate  (\ref{eq20}) as
\be
\begin{bmatrix}
    {\text{Tr}(\mathbf{\dot{A}}^H(\theta)\mathbf{\dot{A}}(\theta)\mathbf{R}_x)-\Gamma}&{\text{Tr}(\mathbf{\dot{A}}^H(\theta)\mathbf{A}(\theta)\mathbf{R}_x)}\\
    {\text{Tr}(\mathbf{A}^H(\theta)\mathbf{\dot{A}}(\theta)\mathbf{R}_x)}&{\text{Tr}(\mathbf{A}^H(\theta)\mathbf{A}(\theta)\mathbf{R}_x)}
\end{bmatrix}\succeq0\label{eq21}.
\ee
Non-convex $\mathbf{P}_2$ is reformulated as $\mathbf{P}_3$ below. The variables for $\mathbf{P}_3$ are specified in set $A_2= \{\mathbf{W}_k ,\lambda_{1,k},\lambda_{2,k},r_k,\epsilon_k,v_k\}$:
\begin{flalign}
    \mathbf{P}_3:\hspace{1mm} &{\max_{A_2}} \hspace{1mm}\text{EE} =   {\sum_{k=1}^K f_k(2 q_k\sqrt{\text{log}_2(1+r_k)}}-q_k^2{P_k})\label{eq22}\\
    \text{s.t.}
    & \ \text{(\ref{eq14c})},\text{(\ref{eq14e})},\text{(\ref{eq14f})},\text{(\ref{eq18})},\text{(\ref{eq19})},\text{(\ref{eq21})}.\notag
\end{flalign}
$\mathbf{W}_k$ and $r_k$ are obtained by $\mathbf{P}_3$. Note that auxiliary variables $q_k$ and $\mathbf{z}_k$ need to be updated from variables in $ \mathbf{P}_3$, instead of using (\ref{eq12}) and (\ref{eq13}). Specifically, $q_k^*$ becomes
\be
q_k^*=\frac{\sqrt{\text{log}_2(1+r_k)}}{\text{Tr}(\mathbf{W}_k)+P_0/K}\label{q_k_new}.
\ee

Through $\mathbf{P}_3$, it is readily seen that $\mathbf{w}_k^H\mathbf{H}_k^H \mathbf{z}_k = r_k$ holds.  However, $\mathbf{z}_k$ is coupled with $\mathbf{H}_k^H$, which contains unknown channel estimation error. Equivalently, $\mathbf{z}_k$ can be obtained from the following problem.
\begin{subequations}
    \begin{flalign}
        \mathbf{P}_4 :&\max_{\mathbf{z}_k,\lambda_{z,k},t} \hspace{2mm} t\label{p4_zk_eq1}\\
        \text{s.t.}
        & \ \mathbf{w}_k^H\mathbf{H}_k^H \mathbf{z}_k\geq t\label{p4_zk_eq2}, 
        \ \mathbf{w}_k^H\mathbf{H}_k^H \mathbf{z}_k\leq r_k.
    \end{flalign}
\end{subequations}
Here, $t \in \mathbb{R}$ is the introduced auxiliary variable. By applying \emph{S}-procedures on constraints  (\ref{p4_zk_eq2}), $\mathbf{P}_4$ is transformed as $\mathbf{P}_5$:
\begin{subequations}
    \begin{flalign}
        \mathbf{P}_5 :&\max_{\mathbf{z}_k,\lambda_{z,k},t}\hspace{2mm}t \label{zk_eq1}\\
        \text{s.t.}
        &\begin{bmatrix}
        \lambda_{z,k}\mathbf{I}_N&{\mathbf{z}_k}/{2}\\{\mathbf{z}_k^H}/{2}&(\mathbf{w}_k^H\mathbf{\hat{H}}_k^H \mathbf{z}_k)-t-\lambda_{z,k}\phi_k^2 \mathbf{w}_k^H\mathbf{w}_k
        \end{bmatrix}\succeq0,&\label{zk_eq2}\\
        &\begin{bmatrix}
        \lambda_{z,k}\mathbf{I}_N&-{\mathbf{z}_k}/{2}\\-{\mathbf{z}_k^H}/{2}&-(\mathbf{w}_k^H\mathbf{\hat{H}}_k^H \mathbf{z}_k)+r_k-\lambda_{z,k}\phi_k^2 \mathbf{w}_k^H\mathbf{w}_k
    \end{bmatrix}\succeq0, \label{zk_eq3}
    \end{flalign}
\end{subequations}
where $\lambda_{z,k}$ is non-negative variable. 
With known $q_k $ and  $\mathbf{z}_k$, $\mathbf{P}_3$ and $\mathbf{P}_5$ are now in the convex form and can be solved via standard optimization toolbox such as  CVX \cite{grant2014cvx}. The iterative process is summarized in \textbf{Algorithm 1}.
\begin{algorithm}
    \caption{Robust beamforming for EE optimization}\label{alg:cap}
    \begin{algorithmic}
        \Require  Initialize parameters, set the counter $j=0 $ and the convergence precision $p_\text{con}$,
        \While{$|\text{EE}_j-\text{EE}_{j-1}|\geq p_\text{con}$}
            \State\textbf{Step 1} $j=j+1$,
            \State\textbf{Step 2}  Update $\mathbf{W}_k$ and $\text{EE}_{j}$ by solving $\mathbf{P}_3$ with fixed \\$\mathbf{z}_k$ and $q_k$,
            \State\textbf{Step 3} Apply EVD on $\mathbf{W}_k$ to update $\mathbf{w}_k$,
            \State\textbf{Step 4} Update $q_k^*$, by  (\ref{q_k_new}),
            \State\textbf{Step 5} Update each $\mathbf{z}_k^*$ from $\mathbf{P}_5$.
        \EndWhile
    \end{algorithmic}
\end{algorithm}
\subsection{Rank Analysis}
To investigate the impact of SDR, we perform the rank analysis of $\mathbf{W}_k$. Specifically, we apply the KKT conditions on  (\ref{eq22}) to obtain the rank of its optimal result.  To match the dimensions of constraints, we set $A_3=\{G_k\in\mathbb{R},O_k\in\mathbb{R},E_k\in\mathbb{R},\mathbf{J}_k\in\mathbb{C}^{(M_t+1)\times (M_t+1)},\mathbf{U}_k\in\mathbb{C}^{(M_t+1)\times (M_t+1)},\mathbf{T}\in \mathbb{C}^{2\times 2},\mathbf{Y}_k\in\mathbb{C}^{M_t\times M_t}\}$ to denote the Lagrangian dual variables of the constraints in $\mathbf{P}_3$ and $\mathbf{W}_k \succeq 0$, respectively. All matrix variables are positive semidefinite.

Hence, the Lagrange dual function for  (\ref{eq22}) is 
\begin{flalign}
    &\mathcal{L}(A_2,A_3)=\sum_{k=1}^K f_k(2q_k\sqrt{ \text{log}_2(1+r_k)}-q_k^2\text{Tr}(\mathbf{W}_k))\notag
\end{flalign}
\begin{flalign}
    &-\text{Tr}(\mathbf{T}(\begin{bmatrix}
        {\text{Tr}(\mathbf{\dot{A}}^H(\theta)\mathbf{\dot{A}}(\theta)\mathbf{R}_x)-\Gamma}&{\text{Tr}(\mathbf{\dot{A}}^H(\theta)\mathbf{A}(\theta)\mathbf{R}_x)}\\
        {\text{Tr}(\mathbf{A}^H(\theta)\mathbf{\dot{A}}(\theta)\mathbf{R}_x)}&{\text{Tr}(\mathbf{A}^H(\theta)\mathbf{A}(\theta)\mathbf{R}_x)}
        \end{bmatrix}))\notag\\
        &+\sum_{ k = 1 }^K\text{Tr}(G_k(\text{Tr}(\mathbf{W}_k)-P^\text{max}))\notag\\
        &-\sum_{ k = 1 }^K\text{Tr}(E_k(2\sqrt{\text{Re}\{\epsilon_k\}}-v_k-r_k)) \notag\\
    &-\sum_{ k = 1 }^K\text{Tr}(\mathbf{J}_k\begin{bmatrix}
        {\mathbf{W}_k+{\lambda}_{1,k} \mathbf{I}_N}&{\mathbf{W}_k^H \hat{\mathbf{H}}_k^H \mathbf{z}_k}\\ \notag
        {\mathbf{z}_k^H\hat{\mathbf{H}}_k \mathbf{W}_k }&{\text{B}-{\lambda}_{1,k} \phi_k^2\mathbf{z}_k^H  \mathbf{z}_k}
    \end{bmatrix}) \notag\\
&-\sum_{ k = 1 }^K\text{Tr}(\mathbf{U}_k\begin{bmatrix}
    {-\mathbf{\Lambda}_k+{\lambda}_{2,k} \mathbf{I}_N}&{-\mathbf{\Lambda}_k^H \hat{\mathbf{H}}_k^H \mathbf{z}_k}\\
    {-\mathbf{z}_k^H \hat{\mathbf{H}}_k \mathbf{\Lambda}_k }&{\text{C}-{\lambda}_{2,k} \phi_k^2\mathbf{z}_k^H  \mathbf{z}_k}
\end{bmatrix})\notag\\
    &-\sum_{ k = 1 }^K\text{Tr}(\mathbf{Y}_k \mathbf{W}_k)-\sum_{ k = 1 }^KO_k(r_k- \zeta_k )
\end{flalign}
Then the complementary condition of $\mathbf{W}_k$ becomes 
\begin{flalign}
\label{eq24}
    \frac{\partial \mathcal{L}}{\partial \mathbf{W}_k}=&-f_k q_k^2 \mathbf{I}_{M_t}-\mathbf{F}-\Bar{\mathbf{H}}_k (\mathbf{J}_k^H+\mathbf{U}_k^H) \Bar{\mathbf{H}}_k^H\notag\\
    &+\sum_{ i = 1 }^K\Bar{\mathbf{H}}_i \mathbf{U}_i^H \Bar{\mathbf{H}}_i^H+G_k\mathbf{I}_{M_t}-\mathbf{Y}_k=0,
\end{flalign}
where $\mathbf{F}\triangleq\partial \mathcal{\mathbf{P}}/\partial \mathbf{\mathbf{W}_k}$, $\Bar{\mathbf{H}}_k=[\mathbf{I}_{M_t}, \hat{\mathbf{H}}_k^H \mathbf{z}_k]$, $\mathbf{P}$ is defined as 
\be
\mathbf{P}=\text{Tr}(\mathbf{T}(\begin{bmatrix}
    {\text{Tr}(\mathbf{\dot{A}}^H(\theta)\mathbf{\dot{A}}(\theta)\mathbf{R}_x)-\Gamma}&{\text{Tr}(\mathbf{\dot{A}}^H(\theta)\mathbf{A}(\theta)\mathbf{R}_x)}\\
    {\text{Tr}(\mathbf{A}^H(\theta)\mathbf{\dot{A}}(\theta)\mathbf{R}_x)}&{\text{Tr}(\mathbf{A}^H(\theta)\mathbf{A}(\theta)\mathbf{R}_x)}
    \end{bmatrix})).
\ee
Since $\mathbf{Y}_k$ is positive semidefinite, we  then have
\begin{flalign}
    \mathbf{Y}_k=&(G_k-f_k q_k^2) \mathbf{I}_{M_t}-\mathbf{F}-\Bar{\mathbf{H}}_k (\mathbf{J}_k^H+\mathbf{U}_k^H) \Bar{\mathbf{H}}_k^H\notag\\
    &+\sum_{ i = 1 }^K\Bar{\mathbf{H}}_i \mathbf{U}_i^H \Bar{\mathbf{H}}_i^H
    \succeq 0\label{eq25}.
\end{flalign} 
From constraints  (\ref{eq18}) and  (\ref{eq19}), we have 
\begin{flalign}
    (\lambda_{1,k}\mathbf{I}_{M_t}+\mathbf{W}_k)\mathbf{\Bar{H}}_k \mathbf{J}_k \mathbf{\Bar{H}}_k^H&=\lambda_{1,k}[\mathbf{0}_{M_t}, \mathbf{\Bar{H}}_k^H \mathbf{z}_k]\mathbf{J}_k\mathbf{\Bar{H}}_k^H\label{rank1J},\\
    (\lambda_{2,k}\mathbf{I}_{M_t}-\mathbf{\Lambda}_k)\mathbf{\Bar{H}}_k \mathbf{U}_k \mathbf{\Bar{H}}_k^H&=\lambda_{2,k}[\mathbf{0}_{M_t}, \mathbf{\Bar{H}}_k^H \mathbf{z}_k]\mathbf{U}_k\mathbf{\Bar{H}}_k^H\label{rank1U}.
\end{flalign}


Therefore, $\text{rank}(\mathbf{\Bar{H}}_k \mathbf{J}_k \mathbf{\Bar{H}}_k^H) =1$, $\text{rank}(\mathbf{\Bar{H}}_k \mathbf{U}_k \mathbf{\Bar{H}}_k^H)=1$ \cite{7922522}. Then, adding  (\ref{rank1J}) and  (\ref{rank1U}) gives
\begin{flalign}
    &\lambda_{1,k}\mathbf{\Bar{H}}_k \mathbf{J}_k \mathbf{\Bar{H}}_k^H+\mathbf{W}_k\mathbf{\Bar{H}}_k \mathbf{J}_k \mathbf{\Bar{H}}_k^H+\lambda_{2,k}\mathbf{\Bar{H}}_k \mathbf{U}_k \mathbf{\Bar{H}}_k^H-\notag\\
    &\mathbf{\Lambda}_k\mathbf{\Bar{H}}_k \mathbf{U}_k \mathbf{\Bar{H}}_k^H=
    [\mathbf{0}_{M_t}, \mathbf{\Bar{H}}_k^H \mathbf{z}_k](\lambda_{1,k}\mathbf{J}_k\mathbf{\Bar{H}}_k^H+\lambda_{2,k}\mathbf{U}_k\mathbf{\Bar{H}}_k^H)\label{rank1sum},
\end{flalign}
Since $\text{rank}(AB) \leq \min(\text{rank}(A), \text{rank}(B))$, therefore, it is easy to observe that the right-handed side of (\ref{rank1sum}) is rank-1  and each item at the left-handed is rank-1 as well. $\mathbf{\Bar{H}}_k \mathbf{J}_k \mathbf{\Bar{H}}_k^H$ and $\mathbf{\Bar{H}}_k \mathbf{U}_k \mathbf{\Bar{H}}_k^H$ are linearly dependent. We combine Lagrangian dual variables semidefinite conditions $\mathbf{J}_k \succeq 0$ and $\mathbf{U}_k \succeq 0$
\begin{flalign}
    \mathbf{\Bar{H}}_k \mathbf{J}_k \mathbf{\Bar{H}}_k^H&=\mathbf{j}_k\mathbf{j}_k^H, \\
    \mathbf{\Bar{H}}_k \mathbf{U}_k \mathbf{\Bar{H}}_k^H&=\mathbf{u}_k\mathbf{u}_k^H,
\end{flalign}
where $\mathbf{j}_k=\delta_k\mathbf{u}_k$, $\mathbf{j}_k, \mathbf{u}_k \in \mathbb{C}^{M_t\times 1}$, $\delta_k$ is a linear scaling factor. Therefore,  $\mathbf{\Bar{H}}_k \mathbf{J}_k \mathbf{\Bar{H}}_k^H=\delta_k^2\mathbf{\Bar{H}}_k \mathbf{U}_k \mathbf{\Bar{H}}_k^H$. We reformulate  (\ref{eq25}) as
\begin{flalign}
    \mathbf{Y}_k=(G_k-f_k q_k^2) \mathbf{I}_{M_t}-\mathbf{F}-(1+\delta_k^2)\mathbf{u}_k\mathbf{u}_k^H+\sum_{i=1}^{K}\mathbf{u}_i\mathbf{u}_i^H
    \succeq 0.
\end{flalign}
With similar steps in \cite{9652071}, $\text{rank}(\mathbf{Y}_k)=\mathbf{M}_t-1$. Since KKT condition ensures $\mathbf{W}_k\mathbf{Y}_k=0$,  $\text{rank}(\mathbf{W}_k)=1$ is proved.

\section{Simulation Results}
In this section, we provide the simulation results for validating the proposed iterative EE optimization algorithm. 
Unless otherwise stated, we set the ISAC system with $K=3$ users, each with $N=2$ receiving antennas, while the weight $f_k$ for each user are set  to be $f_1 = 0.3, f_2 = 0.35, f_3 = 0.35$, respectively. The BS antenna is configured as $M_t=6,8,10,12,14$, and $M_r=M_t+2$ in the following experiments. We assume only one target that needs to be detected in the direction $\theta = \pi/3$. $P^\text{max}=21$ dBm, $ P_0=30$ dBm, $\zeta_k=0.01$. The estimated communication channel follows Rayleigh distribution with path-loss of $-120$ dB, and the background noise level is $\sigma_m=-100$ dBm. For the sensing part, the path-loss is $-120$ dB and background noise level is $\sigma_s=-90$ dBm.  The signal frame length is $L = 30$. 

\begin{figure}[!h]
    \centering
    \includegraphics[scale=0.36]{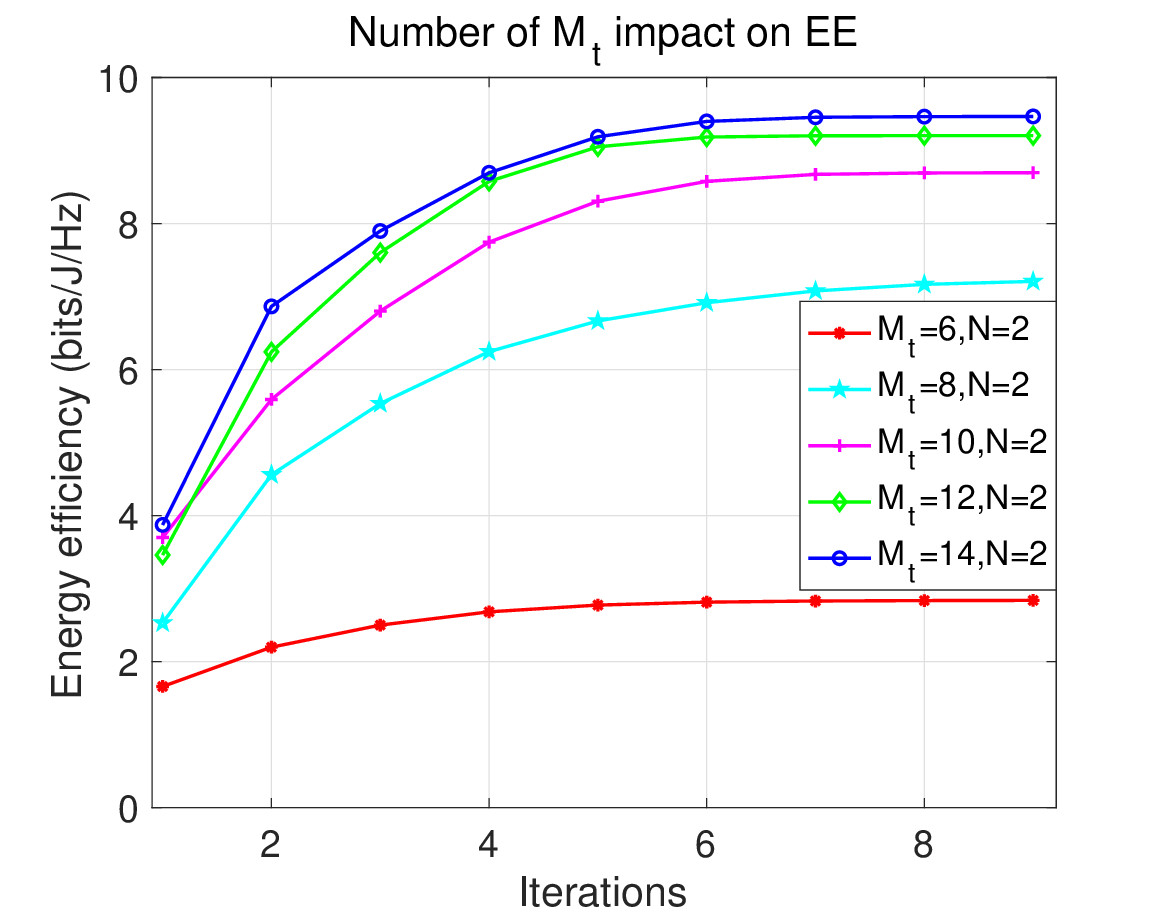}
    \caption{Convergence of the proposed algorithm}
    \label{result1}
\end{figure}

We first demonstrate the  convergence performance of the proposed algorithm in Fig. \ref{result1}. The channel uncertainty factor and CRB requirement are $\phi=0.1$, $ \sqrt{\rho}=0.0033$ (\text{root-CRB}). It can be readily seen that the proposed algorithm has a good convergence performance, usually takes less than 6 iterations to converge. Besides, the system EE increases along with the number of transmitter antennas, which aligns with the intuition of benefits from extra degree-of-freedom (DoF) provided by MIMO systems. 

\begin{figure}[!h]
    \centering
    \includegraphics[scale=0.36]{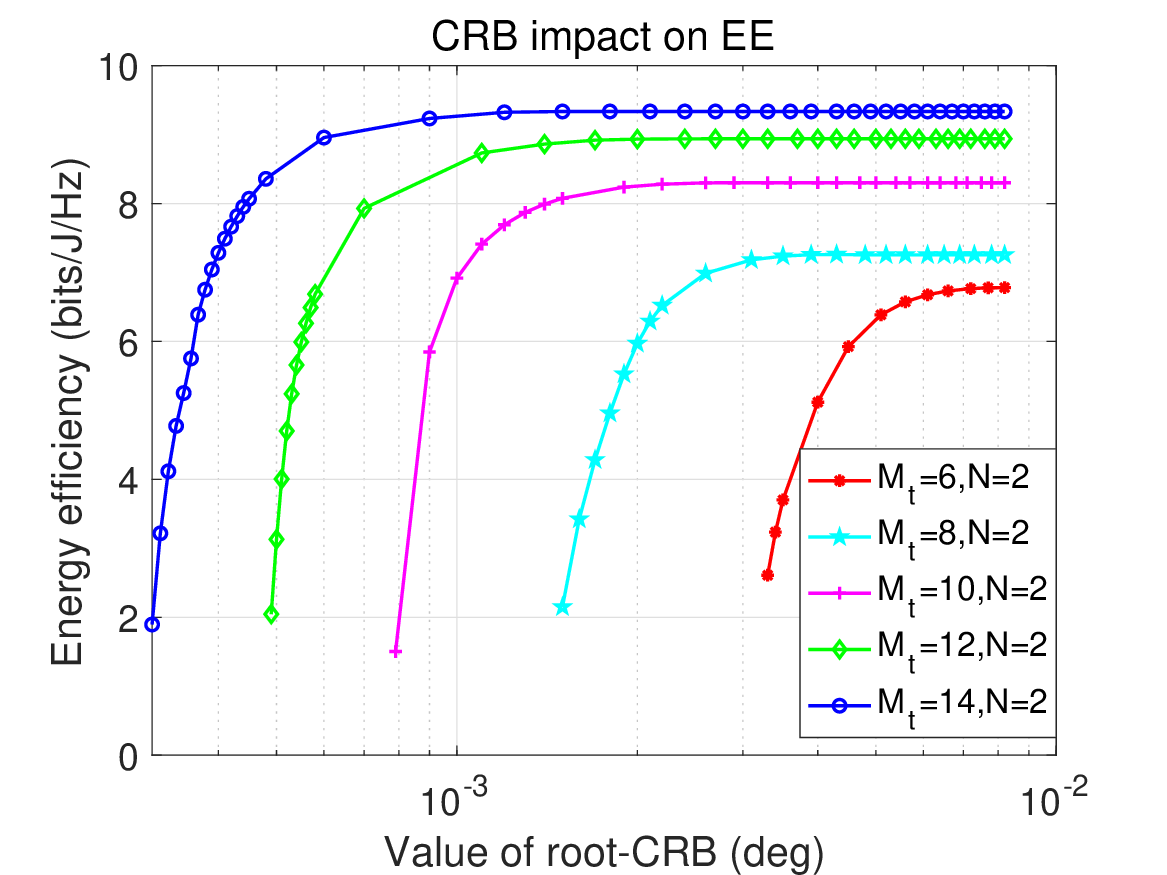}
    \caption{Impact of root-CRB  on energy efficiency}
    \label{result2}
\end{figure}
Fig. \ref{result2} shows  the impact of sensing objective (CRB) on the communication objective (system EE), which illustrates  the performance trade-off of two distinct metrics.  We set the channel uncertainty factor $\phi = 0.3$. As expected, a more stringent requirement of sensing (lower root-CRB) leads to a degraded EE performance. This can be explained by the fact that more communication resources, particularly power for beamforming, is steered to ensure sensing functionality. Besides, more transmitter antennas again lead to a higher sensing and communication performance, due to higher DoF. 

\begin{figure}[!h]
    \centering
    \includegraphics[scale=0.36]{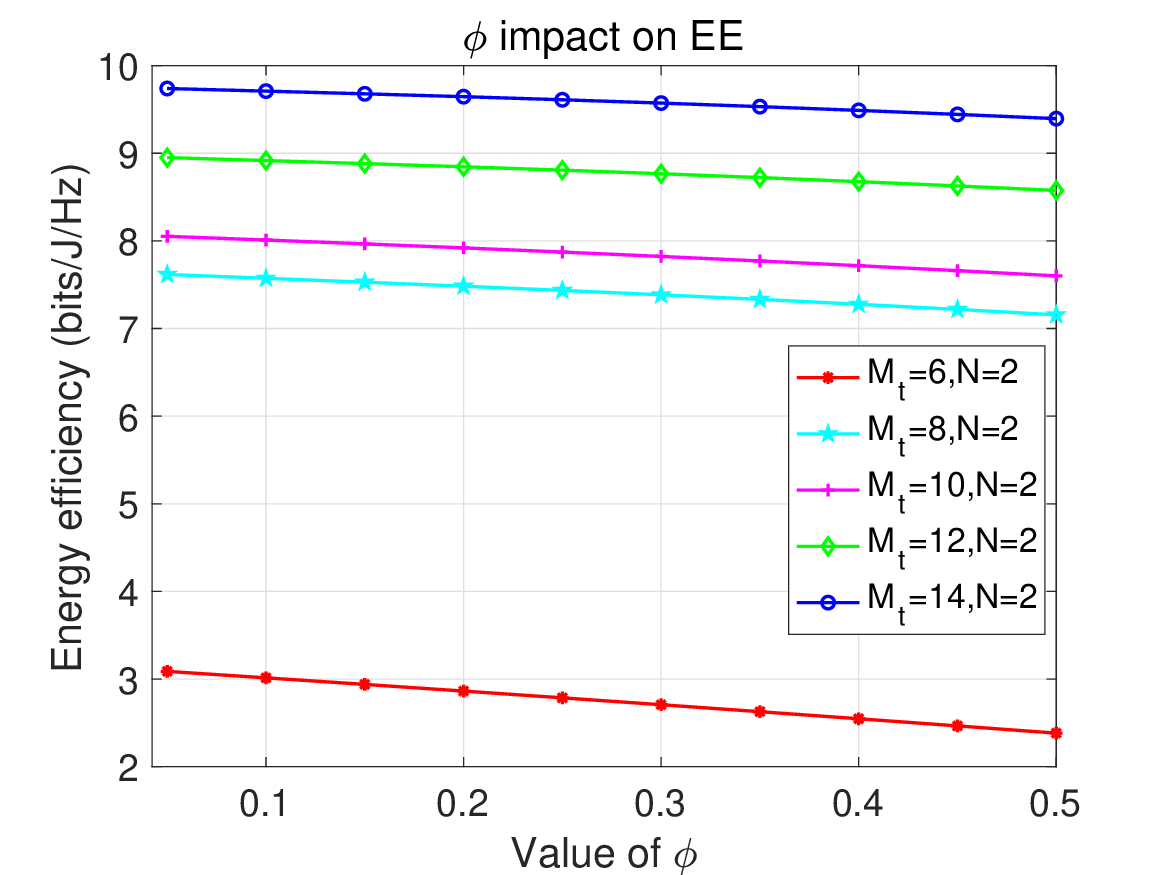}
    \caption{Channel estimation errors impact}
    \label{result3}
\end{figure}

Finally, Fig. \ref{result3} analyzes the impact of channel estimation error. Here, we set the root-CRB $\sqrt{\rho} = 0.0033$ and with channel uncertainty factor $\phi$ in the range of 0.05 to 0.5. Overall, we can observe that system EE decreases with the increase of channel uncertainty. Intuitively, this can be explained with the fact that a higher transmission power is needed to compensate for the channel uncertainty such that SINR constraint for each user is met. A slow decay of system EE, on the other hand, shows the proposed algorithm can effectively tackle channel estimation errors.

\section{Conclusion}
In this paper, we proposed an algorithm to maximize system EE with channel estimation errors in the vehicular ISAC system. To address the channel uncertainty with spectral norm upper bound, we formulated an optimization problem and sequentially transform it into a convex form by SDR, Schur complement, and \emph{S}-procedure. We also provided an in-depth analysis for the rank of optimal solution and proved to be 1 by Lagrangian dual function and KKT conditions. Finally, we conducted numerical experiments to validate the proposed algorithm in terms of its convergence, sensing and communication perforamnce tradeoff, and the impact of channel errors. 
\bibliographystyle{IEEEtran} 
\bibliography{lib}

\begin{thebibliography}{10}
\providecommand{\url}[1]{#1}
\csname url@samestyle\endcsname
\providecommand{\newblock}{\relax}
\providecommand{\bibinfo}[2]{#2}
\providecommand{\BIBentrySTDinterwordspacing}{\spaceskip=0pt\relax}
\providecommand{\BIBentryALTinterwordstretchfactor}{4}
\providecommand{\BIBentryALTinterwordspacing}{\spaceskip=\fontdimen2\font plus
\BIBentryALTinterwordstretchfactor\fontdimen3\font minus
  \fontdimen4\font\relax}
\providecommand{\BIBforeignlanguage}[2]{{%
\expandafter\ifx\csname l@#1\endcsname\relax
\typeout{** WARNING: IEEEtran.bst: No hyphenation pattern has been}%
\typeout{** loaded for the language `#1'. Using the pattern for}%
\typeout{** the default language instead.}%
\else
\language=\csname l@#1\endcsname
\fi
#2}}
\providecommand{\BIBdecl}{\relax}
\BIBdecl

\bibitem{10061429}
X.~Meng, F.~Liu, C.~Masouros, W.~Yuan, Q.~Zhang, and Z.~Feng, ``Vehicular
  connectivity on complex trajectories: Roadway-geometry aware isac
  beam-tracking,'' \emph{IEEE Transactions on Wireless Communications}, pp.
  1--1, 2023.

\bibitem{DSRC_Pc}
J.~B. Kenney, ``Dedicated short-range communications (dsrc) standards in the
  united states,'' \emph{Proceedings of the IEEE}, vol.~99, no.~7, pp.
  1162--1182, 2011.

\bibitem{cv2x}
S.~Gyawali, S.~Xu, Y.~Qian, and R.~Q. Hu, ``Challenges and solutions for
  cellular based v2x communications,'' \emph{IEEE Communications Surveys \&
  Tutorials}, vol.~23, no.~1, pp. 222--255, 2020.

\bibitem{IoTJ_review_vcn}
X.~Cheng, D.~Duan, S.~Gao, and L.~Yang, ``Integrated sensing and communications
  (isac) for vehicular communication networks (vcn),'' \emph{IEEE Internet of
  Things Journal}, vol.~9, no.~23, pp. 23\,441--23\,451, 2022.

\bibitem{Network_V2X_review}
Y.~Zhong, T.~Bi, J.~Wang, J.~Zeng, Y.~Huang, T.~Jiang, Q.~Wu, and S.~Wu,
  ``Empowering the v2x network by integrated sensing and communications:
  Background, design, advances, and opportunities,'' \emph{IEEE Network},
  vol.~36, no.~4, pp. 54--60, 2022.

\bibitem{du2023towards}
Z.~Du, F.~Liu, Y.~Li, W.~Yuan, Y.~Cui, Z.~Zhang, C.~Masouros, and B.~Ai,
  ``Towards isac-empowered vehicular networks: Framework, advances, and
  opportunities,'' \emph{arXiv preprint arXiv:2305.00681}, 2023.

\bibitem{10251151}
Z.~Ren, Y.~Peng, X.~Song, Y.~Fang, L.~Qiu, L.~Liu, D.~W.~K. Ng, and J.~Xu,
  ``Fundamental crb-rate tradeoff in multi-antenna isac systems with
  information multicasting and multi-target sensing,'' \emph{IEEE Transactions
  on Wireless Communications}, pp. 1--1, 2023.

\bibitem{9416177}
X.~Wang, Z.~Fei, Z.~Zheng, and J.~Guo, ``Joint waveform design and passive
  beamforming for ris-assisted dual-functional radar-communication system,''
  \emph{IEEE Transactions on Vehicular Technology}, vol.~70, no.~5, pp.
  5131--5136, 2021.

\bibitem{9652071}
F.~Liu, Y.-F. Liu, A.~Li, C.~Masouros, and Y.~C. Eldar, ``Cramér-rao bound
  optimization for joint radar-communication beamforming,'' \emph{IEEE
  Transactions on Signal Processing}, vol.~70, pp. 240--253, 2022.

\bibitem{9761984}
Z.~He, W.~Xu, H.~Shen, Y.~Huang, and H.~Xiao, ``Energy efficient beamforming
  optimization for integrated sensing and communication,'' \emph{IEEE Wireless
  Communications Letters}, vol.~11, no.~7, pp. 1374--1378, 2022.

\bibitem{zou2023energy}
J.~Zou, S.~Sun, C.~Masouros, Y.~Cui, Y.~Liu, and D.~W.~K. Ng,
  ``Energy-efficient beamforming design for integrated sensing and
  communications systems,'' \emph{arXiv preprint arXiv:2307.04002}, 2023.

\bibitem{li2023isac9652071}
Y.~Li, F.~Liu, Z.~Du, W.~Yuan, and C.~Masouros, ``Isac-enabled v2i networks
  based on 5g nr: How many overheads can be reduced?'' \emph{arXiv preprint
  arXiv:2301.12787}, 2023.

\bibitem{10217169}
H.~Hua, T.~X. Han, and J.~Xu, ``Mimo integrated sensing and communication:
  Crb-rate tradeoff,'' \emph{IEEE Transactions on Wireless Communications}, pp.
  1--1, 2023.

\bibitem{10009894}
J.~Mu, W.~Ouyang, Z.~Jing, B.~Li, and F.~Zhang, ``Energy-efficient interference
  cancellation in integrated sensing and communication scenarios,'' \emph{IEEE
  Transactions on Green Communications and Networking}, vol.~7, no.~1, pp.
  370--378, 2023.

\bibitem{9889690}
T.~Lin and Y.~Zhu, ``Robust precoding design for massive mimo: An efficient
  fractional programming-based approach,'' \emph{IEEE Communications Letters},
  vol.~26, no.~12, pp. 2999--3003, 2022.

\bibitem{8479337}
H.~Sun, F.~Zhou, R.~Q. Hu, and L.~Hanzo, ``Robust beamforming design in a noma
  cognitive radio network relying on swipt,'' \emph{IEEE Journal on Selected
  Areas in Communications}, vol.~37, no.~1, pp. 142--155, 2019.

\bibitem{7922522}
F.~Alavi, K.~Cumanan, Z.~Ding, and A.~G. Burr, ``Robust beamforming techniques
  for non-orthogonal multiple access systems with bounded channel
  uncertainties,'' \emph{IEEE Communications Letters}, vol.~21, no.~9, pp.
  2033--2036, 2017.

\bibitem{spectral_norm_research}
S.~S. Kozat and A.~T. Erdogan, ``Competitive linear estimation under model
  uncertainties,'' \emph{IEEE Transactions on Signal Processing}, vol.~58,
  no.~4, pp. 2388--2393, 2010.

\bibitem{8314727}
K.~Shen and W.~Yu, ``Fractional programming for communication systems—part i:
  Power control and beamforming,'' \emph{IEEE Transactions on Signal
  Processing}, vol.~66, no.~10, pp. 2616--2630, 2018.

\bibitem{grant2014cvx}
M.~Grant and S.~Boyd, ``Cvx: Matlab software for disciplined convex
  programming, version 2.1,'' 2014.

\end{thebibliography}

\end{document}